# Direct comparison of current-induced spin polarization in topological insulator $Bi_2Se_3$ and InAs Rashba states

C. H. Li[1*], O. M. J. van 't Erve[1], S. Rajput[2], L. Li[2], and B. T. Jonker[1]

[1]Materials Science and Technology Division, *Naval Research Laboratory*, Washington, DC 20375 USA.
[2] Department of Physics, *University of Wisconsin*, Milwaukee, WI 53211 USA.

Three-dimensional topological insulators (TIs) exhibit time-reversal symmetry protected, linearly dispersing Dirac surface states. Band bending at the TI surface may also lead to coexisting trivial two-dimensional electron gas (2DEG) states with parabolic energy dispersion that exist as spin-split pairs due to Rashba spin-orbit coupling (SOC). A bias current is expected to generate spin polarization in both systems arising from their helical spin-momentum locking. However, their induced spin polarization is expected to be different in both magnitude and sign. Here, we compare spin potentiometric measurements of bias current-generated spin polarization in $Bi_2Se_3$(111) films where Dirac surface states coexist with trivial 2DEG states, with identical measurements on InAs(001) samples where only trivial 2DEG states are present. We observe spin polarization arising from spin-momentum locking in both cases, with opposite signs of the spin voltage. We present a model based on spin dependent electrochemical potentials to directly derive the signs expected for the TI surface states, and unambiguously show that the dominant contribution to the current-generated spin polarization measured in the TI is from the Dirac surface states. This direct electrical access of the helical spin texture of Dirac and Rashba 2DEG states is an enabling step towards the electrical manipulation of spins in next generation TI and SOC based quantum devices.

* Email: connie.li@nrl.navy.mil

The quest for the realization of efficient generation and electrical control of spin has motivated the search for materials that exhibit strong spin splitting of their electronic states [1,2]. A successful platform has been the two-dimensional electron gas (2DEG) in semiconductor heterostructures, where structural inversion asymmetry along the growth direction lifts spin degeneracy via spin-orbit coupling [3]. In 2DEGs with a parabolic energy dispersion, the Rashba form of spin-orbit coupling leads to a pair of Fermi surfaces that exhibit counter-rotating chiral spin texture, locking spin to the linear momentum (Fig. 1a) [2,4-9]. The further demonstration of electrical gate control of the strength of such spin splitting has led to prospects for prototypical semiconductor field effect spintronic devices [8,9].

Helical spin-momentum locking is also exhibited in a recently discovered quantum phase of matter, three-dimensional topological insulators (TIs), where linearly dispersing metallic surface states populated by massless Dirac fermions coexist with a semiconducting bulk [11-15]. The 2D surface states are occupied by a single spin, and topologically protected by time reversal symmetry, making them robust against scattering, as illustrated in Fig. 1b. This helical spin texture has been observed by spin- and angle-resolved photoemission spectroscopy (ARPES) [16-18]. These measurements have also shown that trivial Rashba spin-split 2DEG states may coexist with the Dirac surface states in TI materials that exhibit surface carrier accumulation arising from band bending [19,20]. The trivial 2DEG states are nested within the linear dispersing Dirac states [19,20] as illustrated in Fig. 1c such that the spin orientation of the higher $k$ 2DEG state is opposite that of the nearest Dirac state.

An unpolarized bias current is predicted to create a net spin polarization due to spin-momentum locking for both the topologically protected TI surface states [21-23] and the Rashba 2DEG states [3-5]. The spin helicities of the two have been shown in momentum-resolved measurements such as spin-ARPES to be opposite [19]. In transport measurements, however, the measured spin-polarization is momentum integrated, and both Dirac and Rashba 2DEG states can contribute to the spin voltage measured at the detector contact. Calculations treating a model $Bi_2Se_3$ surface in which these states coexist found that the spin polarization and sign of the corresponding spin voltage measured at the detector contact is indeed opposite for the TI Dirac state and Rashba 2DEG contributions [25,26]. These calculations were performed for a three-terminal potentiometric geometry for both ballistic and diffusive regimes, using a spin-orbit coupling coefficient of $\alpha = 0.79$ Å obtained from ARPES measurements on $Bi_2Se_3$ [20]. Furthermore, because the Rashba 2DEG states exists as spin-split pairs with Fermi level momenta $k_1 < k_2$, their spin contribution given by *$(k_2–k_1)/(k_2+k_1)$* mostly cancels, and the net spin polarization was shown to be dominated by the TI surface states [14,23,25].

We recently demonstrated the first direct electrical detection of spin-polarization resulting from spin-momentum locking in TI surface states in $Bi_2Se_3$ using spin potentiometric measurements, where the projection of the current-generated spin onto the magnetization of a ferromagnetic/tunnel barrier detector contact was measured as a voltage (Fig. 1d) [24]. Similar investigations on various TI materials using similar FM/tunnel contacts [27-31] have subsequently been reported. However, inconsistent results were reported regarding the sign of the spin signal, perhaps due to variations in material quality, measurement and analysis, or the potential coexistence of the two spin systems. Lee et al [30] measure a spin-voltage consistent with ours [24] in electrically gated $(BiSb)_2Te_3$ samples as they systematically move the Fermi energy from the conduction band edge through the Dirac point to the valence band edge, while others report a spin voltage of opposite sign with a markedly different temperature dependence [29]. This underscores the need to independently probe and compare the characteristics of the Dirac and Rashba band systems.

In this work, we report a direct comparison of the current induced spin polarization measured using identical $Fe/Al_2O_3$ tunneling spin-potentiometric contacts and measurement geometries in two prototype systems, the TI $Bi_2Se_3$, where both Dirac surface states and a Rashba 2DEG are known to coexist, and InAs which exhibits only the Rashba surface 2DEG. We show that the sign of the spin signal measured in the $Bi_2Se_3$ and InAs samples is indeed opposite, and the temperature dependence is markedly different. We further develop a model based on spin-dependent electrochemical potentials to explicitly illustrate the measurement and derive the sign of the spin voltage expected for the TI surface states, which corroborates our experimental observation here and in previous work [24]. These results unambiguously show that the dominant contribution to the bias current-generated spin voltage detected by the magnetic tunnel barrier spin potentiometric measurement in the TI is from the Dirac surface states.

**Results**

**Current generated spin in TI Dirac states.** $Bi_2Se_3$ samples were grown by molecular beam epitaxy (MBE) on $Al_2O_3(0001)$ substrates using a two-step process (see Methods). $Fe/Al_2O_3$ spin detector contacts were deposited on both $Bi_2Se_3$ and InAs samples in the same MBE system. The samples were then processed into identical device structures illustrated in Figs. 2a & b. For the spin potentiometric measurements, a bias current is applied between the two Ti/Au current leads on either end of the device mesa, and a voltage is measured between the pairs of ferromagnetic (red) detector and corresponding non-magnetic Au/Ti (yellow) reference contacts.

When an unpolarized current flows between the two outer Ti/Au contacts, a spontaneous spin polarization is produced in the $Bi_2Se_3$ surface states throughout the channel due to spin-momentum locking. The projection of this spin onto the magnetization of the ferromagnetic detector contact is recorded as a voltage with a high-impedance voltmeter (>1 Giga-ohm). An in-plane magnetic field is applied to rotate the magnetization of the Fe detector contact, so that the projection of the current-generated spins onto the detector magnetization changes, resulting in a change in sign and magnitude of the detector voltage. Here we define the positive current to be holes flowing from left to right along the $+x$ axis, and the positive magnetic field to be pointing in the $+y$ direction.

The detector voltage as a function of magnetic field for +2 mA bias current at T = 10 K is shown in Figure 2c, after a simple linear background subtraction [24] and centering around the vertical axis. Electron flow in the –x direction generates a spin in the +y direction due to spin-momentum locking in the TI Dirac states. At positive magnetic field > 60 Oe, when the detector magnetization is saturated and completely parallel to the spin direction, a constant low voltage is observed. As the magnetic field decreases to small negative values around -50 Oe (coercive field of the magnetic contact) (red trace), an abrupt increase in detector voltage is seen as the detector magnetization reverses to be antiparallel with the TI surface-state spin, and the overall scan exhibits a single step-like behavior. When the field sweep direction is reversed to increase from negative to positive values (black trace), the detector voltage is constant until reaching the positive coercive field of 50 Oe, where the voltage abruptly decreases as the detector magnetization again switches and becomes parallel with the spin orientation of the TI surface state.

Changing the current direction to -2 mA (Fig. 2d), i.e., electrons flowing from left to right in the +x direction generate a spin in the –y direction. At positive magnetic field above saturation, the detector magnetization is antiparallel to the current-generated spin, and a constant high voltage is observed, while at negative magnetic field where the detector magnetization is parallel to the spin, a constant low voltage is seen. Comparing figures 2c and 2d, the hysteresis loop simply inverts around the x-axis. This behavior is very reproducible for temperatures up to 250 K, as is shown in Figs. 2e & f for both currents where clear hysteresis curves can still be seen.

**Temperature and bias dependence.** The temperature dependence of the magnitude of the spin signal, $|V(+M) - V(-M)|$, measured at ±2 mA is shown in Fig. 3a. It decreases monotonically with increasing temperature to 175 K, exhibits a small increase between 175 and 250 K, and then disappears at 275 K. The small increase in the spin voltage in the range 175-250 K, and its abrupt suppression by 275 K are not well understood at present. This temperature dependence is similar to our previous observations of $Bi_2Se_3$ on graphene/n-SiC substrates [24], although in our previous work the spin-voltage was observed only up to 150 K. The higher temperature achieved in this case for $Bi_2Se_3$ films grown directly on an insulating $Al_2O_3$ substrate could be attributed to the fact that there is no current shunting through the epi-graphene/SiC conductive substrate used previously, so that a higher fraction of the bias current flows in the TI surface layer to produce the measured spin polarization.

The dependence of the spin signal measured at the detector contact ($\Delta V = V(+M) - V(-M)$) as a function of the bias current at T = 10 K is shown in Fig. 3b, where a nearly linear dependence is observed. This linear behavior of the spin signal with bias current is consistent with model calculations [25], where the voltages measured on the FM detector $V(M)$ were directly related to the bias current and spin polarization by $[V(+M) - V(-M)] = I_b\, R_B\, P_{FM}\, (\boldsymbol{p} \cdot \boldsymbol{M_u})$, (bold case denotes a vector). Here $I_b$ is the (hole) bias current in the $+x$ direction, $R_B$ is the ballistic resistance of the channel, and $P_{FM}$ is the transport spin polarization of the FM detector metal. $\boldsymbol{M_u}$ is a unit vector along the detector magnetization $\boldsymbol{M}$, and $\boldsymbol{p}$ is the degree of spin polarization induced per unit current by both spin-momentum locking in TI Dirac surface states and Rashba spin-orbit coupling in the 2DEG. From the spin signal we measure (e.g., Fig. 3b), assuming that the bias current is shunted equally by each quintuple layer of the $Bi_2Se_3$ film [24], and taking $P_{FM}(Fe) \sim 0.4$, and $k_F \sim 0.15$ Å$^{-1}$, we estimate $\boldsymbol{p} \sim -0.15$, with a sign that's indicative of the TI Dirac states [25].

**Current generated spin in InAs Rashba states.** To further distinguish the sign of the spin signal measured for the TI Dirac surface states from that of potential trivial 2DEG states, we performed similar measurements on InAs(001) samples where only the surface 2DEG states are known to exist [32-36]. It is well known that the downward band bending of the conduction band at the InAs(001) surface leads to an electron accumulation layer and the formation of a surface 2DEG [32-36] (Fig. 4a) that extends ~ 20 nm into the sample [36]. The InAs samples were processed to produce the same

Fe/Al$_2$O$_3$ contact geometry used for the Bi$_2$Se$_3$ measurements (see Materials and Methods). As noted earlier, the Rashba spin-orbit induced polarization is predicted to exhibit the opposite sign to that of TI Dirac states [25] for a given bias current.

The spin-voltage transport data for these InAs samples are shown in Fig. 4b and c. The measurement procedures were identical to those used for the Bi$_2$Se$_3$ samples. Similar hysteresis loops are observed where a constant "high" and "low" voltage is measured when the detector magnetization is fully aligned with the applied field. However, for a given current/electron flow direction, the hysteresis loop is clearly inverted about the horizontal axis relative to that observed for the Bi$_2$Se$_3$ samples (Figs. 2c-f): for a positive bias current, a "high" voltage signal is seen at positive fields above the coercive field of the Fe contact, and a "low" voltage is observed negative fields. Given that the spin detecting contacts (Fe/Al$_2$O$_3$) and contact geometry are the same, and are sensitive only to the orientation of the induced spin polarization regardless of the source, this observation indicates that the bias current-induced spin polarization due to spin-momentum locking in the InAs 2DEG is opposite to that of the Bi$_2$Se$_3$ TI Dirac states, consistent with the theory [25]. In addition, the spin voltage exhibits a weak temperature dependence, decreasing by only ~ 10% from 10 to 300 K, and persists to at least 300 K, as shown in Fig. 4d, consistent with the metallic nature of the 2DEG. This is markedly different than the strong temperature dependence observed for the TI Dirac surface state (Fig. 3a).

**Discussion**

While both the topologically protected Dirac surface states and the Rashba spin-split 2DEG states of the Bi$_2$Se$_3$ are expected to produce a bias current induced spin polarization due to spin-momentum locking, the current-induced spin density is expected to be substantially larger for the Dirac surface states for several reasons. First, as discussed above, for a given momentum direction, the Rashba surface 2DEG states exist as spin-split pairs of opposite spin orientation, and the net induced spin polarization is proportional to *($k_2$-$k_1$)/($k_2$+$k_1$)* [25], where $k_2 > k_1$ (see Fig. 1a). Consequently, the contributions from these spin-split states tend to cancel. In contrast, the Dirac state has only one spin orientation, and no such cancellation occurs. Second, the induced spin polarization is enhanced by a factor *$v_F$/α*>>1 for the TI Dirac surface states, where *$v_F$* is the Fermi velocity of the TI (on the order of $10^5$ m/s) and *α* the strength of the Rashba spin-orbit coefficient in the 2DEG in units of velocity (on the order of $10^3$ m/s) [14, 20, 23]. The fact that in degenerate $Bi_2Se_3$ samples where both the Dirac and Rashba 2DEG states coexist, the sign of the spin signal we observe corresponds to that of the TI Dirac states, corroborates the prediction and expectation that the signal should be dominated by contributions from the Dirac states.

As noted earlier, there are inconsistencies in the sign of the spin signal [*V(+M) - V(-M)*] reported for nominally identical measurements of the bias current induced spin polarization attributed to the TI Dirac states, even for the same $Bi_2Se_3$ material (and stated conventions for current and magnetic field directions). This is reflected in whether a "low" or "high" voltage signal is observed when the detector magnetization is parallel or antiparallel to the induced spin. In our measurements of the TI Dirac surface states, we observe [*V(+M) - V(-M)*] < 0, i.e. a "low" voltage signal when the detector magnetization and TI spin are parallel, and a "high" signal when antiparallel (Fig. 2 and reference 24), as have other groups [30]. In contrast, others report the opposite behavior [29,31], as well as markedly different temperature dependence for the spin voltage in Bi$_2$Se$_3$ films [31].

Hence to directly derive the sign of the spin voltage that should be expected, we develop a model based on the spin-dependent electrochemical potentials generated and their detection by a ferromagnetic detector contact. We note that similar models have been reported in Ref. 29 & 31, although the gradients and/or the reference of the electrochemical potentials are inconsistent with conventional usage in the spintronics community. To better illustrate and contrast the discrepancies, we construct our diagram using notation similar to that of Ref. 29.

We begin with a simple 3-terminal measurement geometry similar to that of Hong *et al*. [25] shown in Fig. 5a. We define the left contact as the positive terminal, and the right contact as the negative terminal used as the reference contact, as used in our measurements. The positive magnetic field direction (and detector magnetization) is again defined to be in the +y direction, with positive (hole) current flowing in the +x direction. We present a diagram of the electric field and voltage ($V$), where $V$ is directly related to electrochemical potential ($\mu$) by $\mu = -eV$, [37,38] where $e$ is the electron charge (taken to be a positive quantity). With these conventions, the voltage reference point and gradient of the electric field are unambiguously defined. In the following, we discuss the measurement in terms of both the voltage and electrochemical potential.

For a positive current ($I > 0$) (Fig. 5b & c), electron momentum $k_e$ is from right to left in the $-x$ direction, and the voltage of the left electrode ($V_L$) is high relative to that of the right electrode ($V_R$). The right (reference) contact need not be grounded, so we indicate a zero reference $V = \mu = 0$ by the yellow line common to panels 5b and 5c. Thus the gradient of the electric field has a negative slope (Fig. 5b). The profile of the electrochemical potential (Fig. 5c), $\mu = -eV$, is merely the mirror image of the electric field/voltage profile across the $V = \mu = 0$ axis. Here the left electrode (**L**) has a more negative (larger magnitude) electrochemical potential than the right electrode (**R**), and the profile has a positive slope.

When a bias current flows through the TI Dirac surface states, a net spin polarization is created due to spin-momentum locking with a direction determined by the electron momentum: for $k_e$ along the $-x$ direction (positive bias current), the induced spin is oriented along the $+y$ axis and referred to as "spin-up". Consequently, the electrochemical potential splits for spin-up and spin-down electrons, as represented by the blue $\mu_\uparrow$ and red $\mu_\downarrow$ lines in Fig. 5c, where $\mu_\uparrow$ for the spin-up electrons is larger (in magnitude, i.e. more negative) than that of the spin-down $\mu_\downarrow$ (i.e., $|\mu_\uparrow| > |\mu_\downarrow|$). The corresponding levels in the voltage diagram of Fig. 5b are again the mirror image across the horizontal axis.

This spin imbalance is probed by the ferromagnetic detector contact. The magnetization of the ferromagnet aligns with the applied external magnetic field above saturation. However, its magnetic moment is opposite to the orientation of its majority spin [39]. Hence the FM detector with $+M$ magnetization (oriented along $+y$) has its majority spin oriented along $-y$, and will probe the spin-down electrochemical potential ($\mu_\downarrow$, $V_\downarrow$) in the channel. Conversely, the detector with $-M$ magnetization probes the spin-up levels ($\mu_\uparrow$, $V_\uparrow$).

Since the right electrode (**R**) is the reference, when the detector magnetization is saturated at positive magnetic field, its voltage $V(+M)$ due to probing the spin-down electron band is $-eV(+M) = \mu_\downarrow - \mu_R$, or $V(+M) = (\mu_\downarrow - \mu_R)/(-e)$. Similarly, when the detector magnetization is saturated at negative magnetic field, its voltage due to probing the spin-up band is $V(-M) = (\mu_\uparrow - \mu_R)/(-e)$. Since $|\mu_\uparrow| > |\mu_\downarrow|$, this yields a high voltage signal at the negative field, when the magnetization is antiparallel to the TI spin (spin-up), and a low voltage at positive field, when it is parallel to the TI spin, as depicted by the hysteresis loop below the potential diagram in Fig. 5c. Note that a simple linear background subtraction and centering around the vertical axis does not change the relative "high" and "low" signals.

A similar analysis can be made directly in terms of the voltage, as shown in the top panels of Fig. 5b. Electrons flowing from right to left in the –x direction create a net spin-up population oriented along +y, and hence the blue $V_\uparrow$ level is higher (i.e., larger in magnitude) than that of the spin-down $V_\downarrow$ (i.e., $V_\uparrow > V_\downarrow$), analogous to $|\mu_\uparrow| > |\mu_\downarrow|$. With the right electrode (*R*) as the reference, the detector voltage at positive magnetic field (*+B, +M*) probing the spin-down level is $V(+M) = (V_\downarrow - V_R)$, and that at negative magnetic field (*-B, -M*) which probes spin-up is $V(-M) = (V_\uparrow - V_R)$. And since $V_\uparrow > V_\downarrow$, at positive field (when magnetization is parallel to TI spin), a low voltage is expected, and at negative field (magnetization antiparallel to TI spin), a high voltage is expected (or *ΔV* (=*V(+M) - V(-M)*) should be negative), exactly as we observe experimentally in $Bi_2Se_3$.

This exercise can be repeated when the current direction (and hence induced spin direction) is reversed, as shown in Figures 5e & e. We note that discrepancies/mistakes can easily be made when discussing electrochemical potentials with negative values. For example in Fig. 1d of Ref. 29, with a positive current and the left electrode (*L*) as the reference (negative) electrode (Fig. 1b), the voltage of the left electrode should be low, and hence the chemical potential of the left electrode should also be low (in magnitude) relative to that of the right, opposite to that shown.

In summary, we directly compare electrical measurements of current-generated spin polarization due to spin-momentum locking in two complimentary systems: $Bi_2Se_3$ with the potential coexistence of both Dirac and Rashba 2DEG surface states, and InAs with only the Rashba 2DEG surface states. We show that the spin voltages measured for the Dirac and Rashba systems are indeed opposite in sign, as predicted by theory [25]. We further develop a model based on spin-splitting of the electrochemical potential to derive the sign of the spin voltage expected for the Dirac states from a potentiometric measurement using a ferromagnetic contact, further confirming that the spin signals we observe from the $Bi_2Se_3$ are consistent with the Dirac surface states. These results demonstrate conclusively that the current-generated spin polarization in TI Dirac and Rashba 2DEG states are indeed opposite, as expected from their different energy band dispersion, and that in a TI it is dominated by the Dirac surface states. These demonstrations of direct electrical detection of the helical spin texture of Dirac and Rashba states is an enabling step towards electrical manipulation of spins in next generation TI and SOC based quantum devices.

**Acknowledgements:**

The NRL authors gratefully acknowledge support from the NRL Nanoscience Institute and core programs at the Naval Research Laboratory, and from the Office of Naval Research. The authors at UW acknowledge support from the Department of Energy (DE-FG02-07ER46228) at the University of Wisconsin, Milwaukee.

**Author contributions**: C.H.L., L.L. and B.T.J. conceived and designed the experiments. S.R. and L.L. grew the $Bi_2Se_3$ films. C.H.L. fabricated the devices and performed the transport measurements with assistance from O.M.J.E. C.H.L., O.M.J.E. and B.T.J. analyzed the data. C.H.L., L.L. and B.T.J. wrote the paper. All authors discussed the results and commented on the manuscript.

**METHODS**

The growth of $Bi_2Se_3$ films was carried out on $Al_2O_3(0001)$ substrates in an ultrahigh vacuum (UHV) system (base pressure $\sim 1\times 10^{-10}$ Torr) that integrates two MBE chambers and a low temperature (5-300K) scanning tunneling microscope (STM). For the growth of $Bi_2Se_3$, a two-step process is used [40]: 2-3 quintuple layers (QL) of $Bi_2Se_3$ were first deposited at a reduced temperature of 100 °C, and the substrate temperature was then slowly raised to 300 °C where the rest of the film was deposited. Bi and Se were supplied via separate Knudsen cells at 460 and 250 °C, respectively [41]. The samples were then removed to air and transferred to a separate MBE system where $Fe/Al_2O_3$ contacts were deposited as described previously [42] and below.

In the case of InAs, an undoped InAs(001) substrate was heated to 520 °C in an As flux to desorb the oxide. The sample was then cooled to room temperature and transferred under ultra-high vacuum into an interconnected MBE system for the growth of $Fe/Al_2O_3$ (the same system used to deposit $Fe/Al_2O_3$ on $Bi_2Se_3$).

The $Fe/Al_2O_3$ contacts were formed on $Bi_2Se_3$ as follows. A 0.7 nm layer of polycrystalline Al was first deposited by MBE, and then oxidized in 200 Torr $O_2$ for 20 min in the presence of UV light in the load-lock chamber. This step was then repeated for a total $Al_2O_3$ thickness of 2 nm. The sample was then transferred under UHV to an interconnected metals MBE chamber, where 20 nm of polycrystalline Fe was deposited at room temperature from a Knudsen cell.

The samples were processed into the device structures illustrated in Fig. 2a & b to enable transport measurements. Standard photolithography and chemical etching methods were used to define the Fe contacts, which ranged in size from 20x20 $\mu m^2$ to 80x80 $\mu m^2$, with adjacent contact separation ranging from 45 to 200 μm.  Ion milling was used to pattern the $Bi_2Se_3$ mesa.   Large Ti/Au contacts were deposited by lift-off in an electron beam evaporator as non-magnetic reference contacts and bias current leads. The Fe contacts were capped with 10 nm Ti / 100 nm Au, and bond pads for wire bonded electrical connections are further electrically isolated using 100 nm of $Si_3N_4$.

Transport measurements were performed in a closed cycle cryostat equipped with an electromagnet (4-300 K, ±1000 Oe).  An unpolarized bias current was applied through the outer Ti/Au contacts on the opposite ends of the $Bi_2Se_3$ mesa, and the voltage on the detector contact was recorded as a function of the in-plane magnetic field applied orthogonal to the electron bias current direction in the TI.

**FIGURE CAPTIONS**

**Fig. 1. Spin-momentum locking and experimental Concept**. Schematic diagram of the spin-momentum locking textures of the Rashba 2DEG (a), Dirac surface states of TI (b), and coexistence of both (c). (d) Experimental concept of the potentiometric measurement.

**Fig. 2. Device schematic and electrical detection of current-generated spin in TI**. Schematic (a) and top view (b) of contact layout with two parallel rows of collinear detector contacts, top row is ferromagnetic (Fe, red), bottom row is non-magnetic reference (Ti/Au). Magnetic field dependence of the voltage measured at the ferromagnetic detector contact with the magnetization collinear with the induced TI spin for bias currents of +2 mA (c) and -2 mA (d). Similar measurements at 250K at +2 mA (e) and -2 mA (f).

**Fig. 3**. **Temperature and bias dependence of TI spin voltage.** Temperature dependence of the spin voltage at +/-2 mA bias current. (Inset: illustration of how $\Delta V = V(M) – V(-M)$ is determined.) (b) Bias current dependence of the ferromagnetic detector voltage.

**Fig. 4**. **Electrical detection of current-generated spin in InAs 2DEG.** (a) Schematic of InAs surface 2DEG formation. Magnetic field dependence of the voltage measured at the ferromagnetic detector contact with the magnetization collinear with the induced 2DEG spin in InAs for bias currents of +5 mA (b) and -5 mA (c). (d) Temperature dependence of the spin voltage at +/-5 mA bias current.

**Fig. 5**. **Model to derive sign of spin signal expected in TI.** (a) Schematic of a simplified 3-terminal device, and definitions of voltage terminals and magnetic field and magnetization directions. Model of the spin potentiometric measurement probing the current-induced spin polarization due to TI surface states for both (b) positive and (c) negative bias currents, based on the spin-dependent electrochemical potentials.

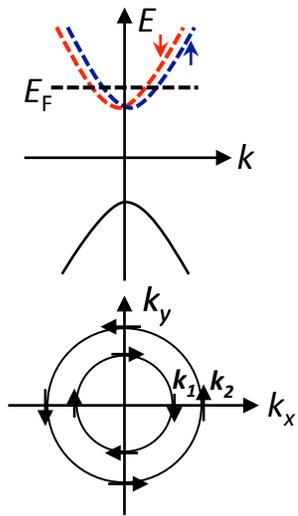
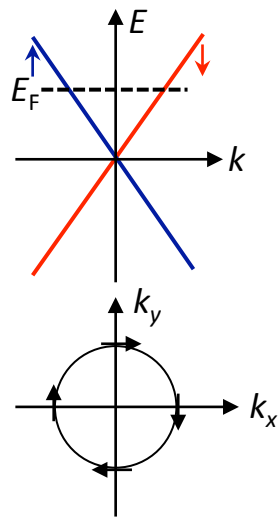
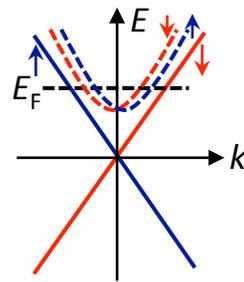
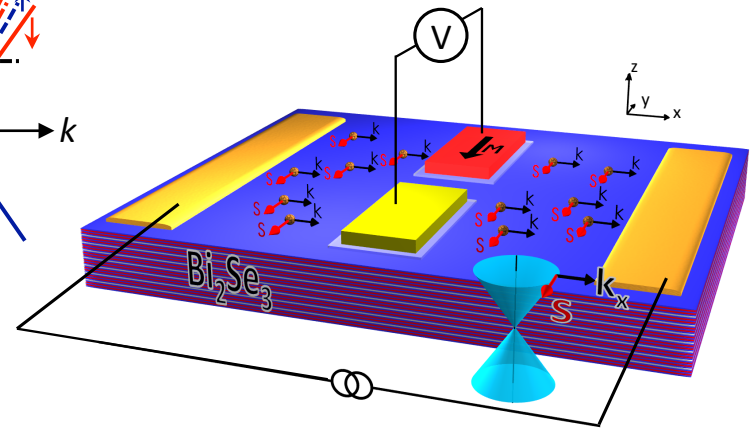

Li et al., Fig. 1

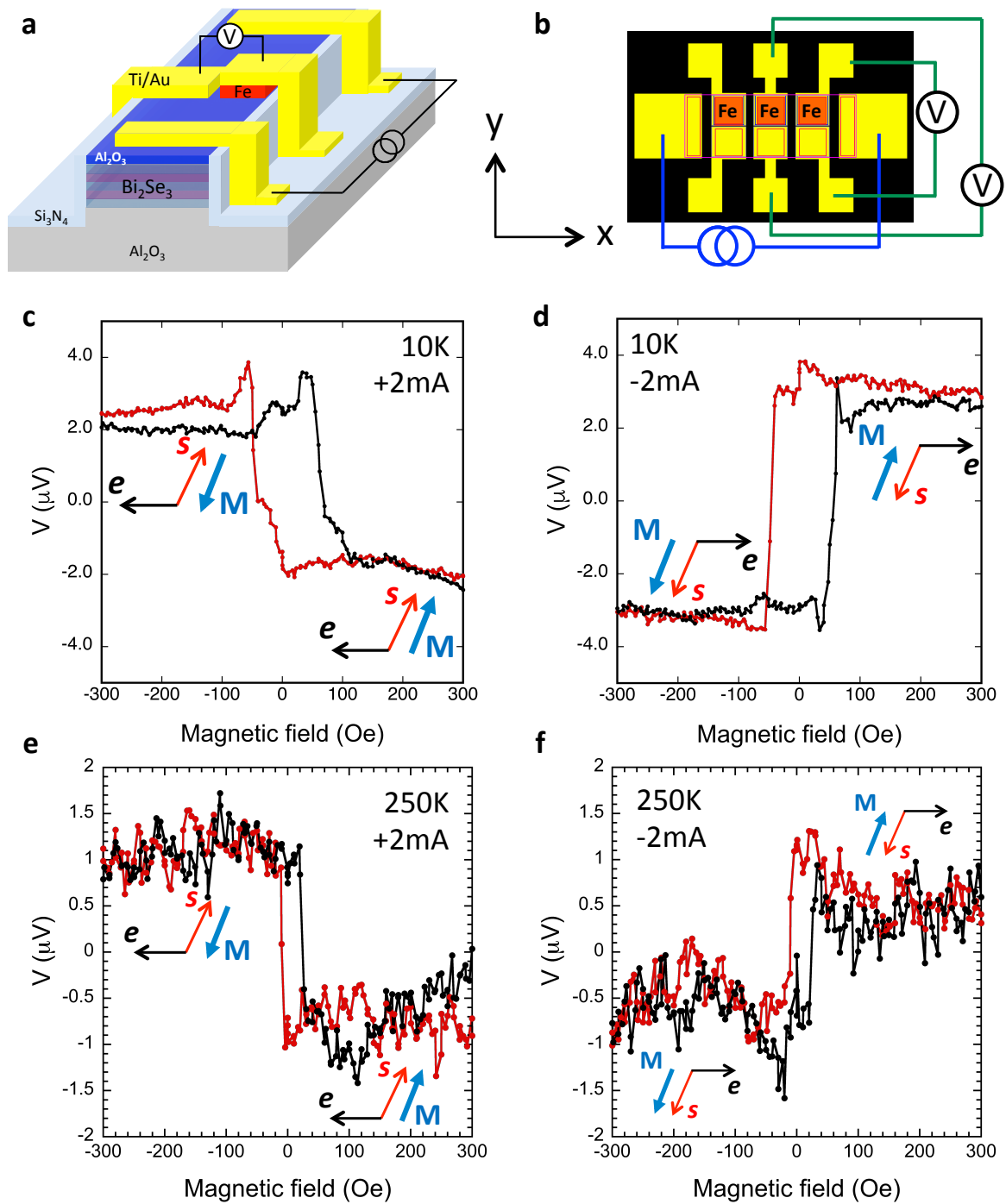

Li et al., Fig. 2

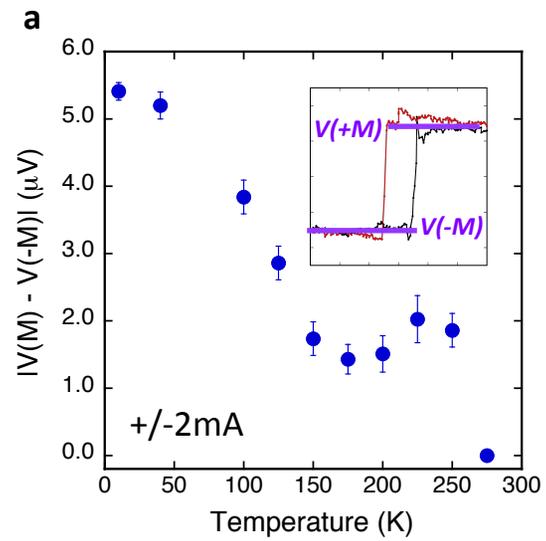

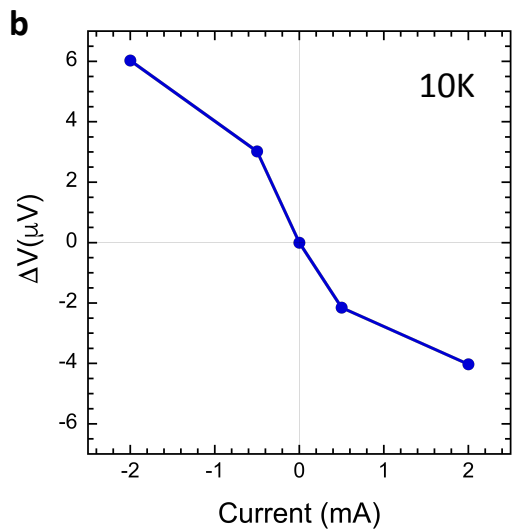

Li et al., Fig. 3

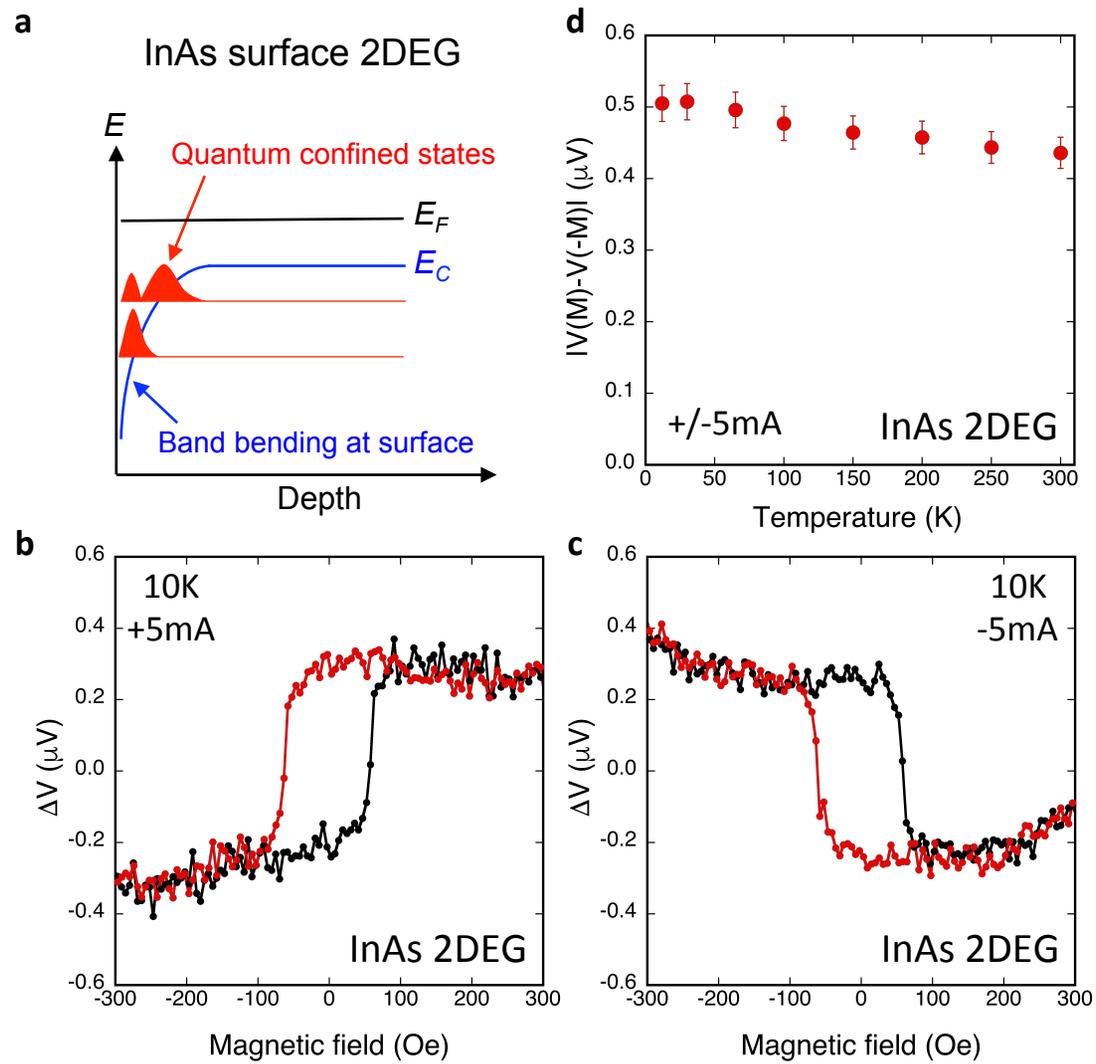

Li et al., Fig. 4

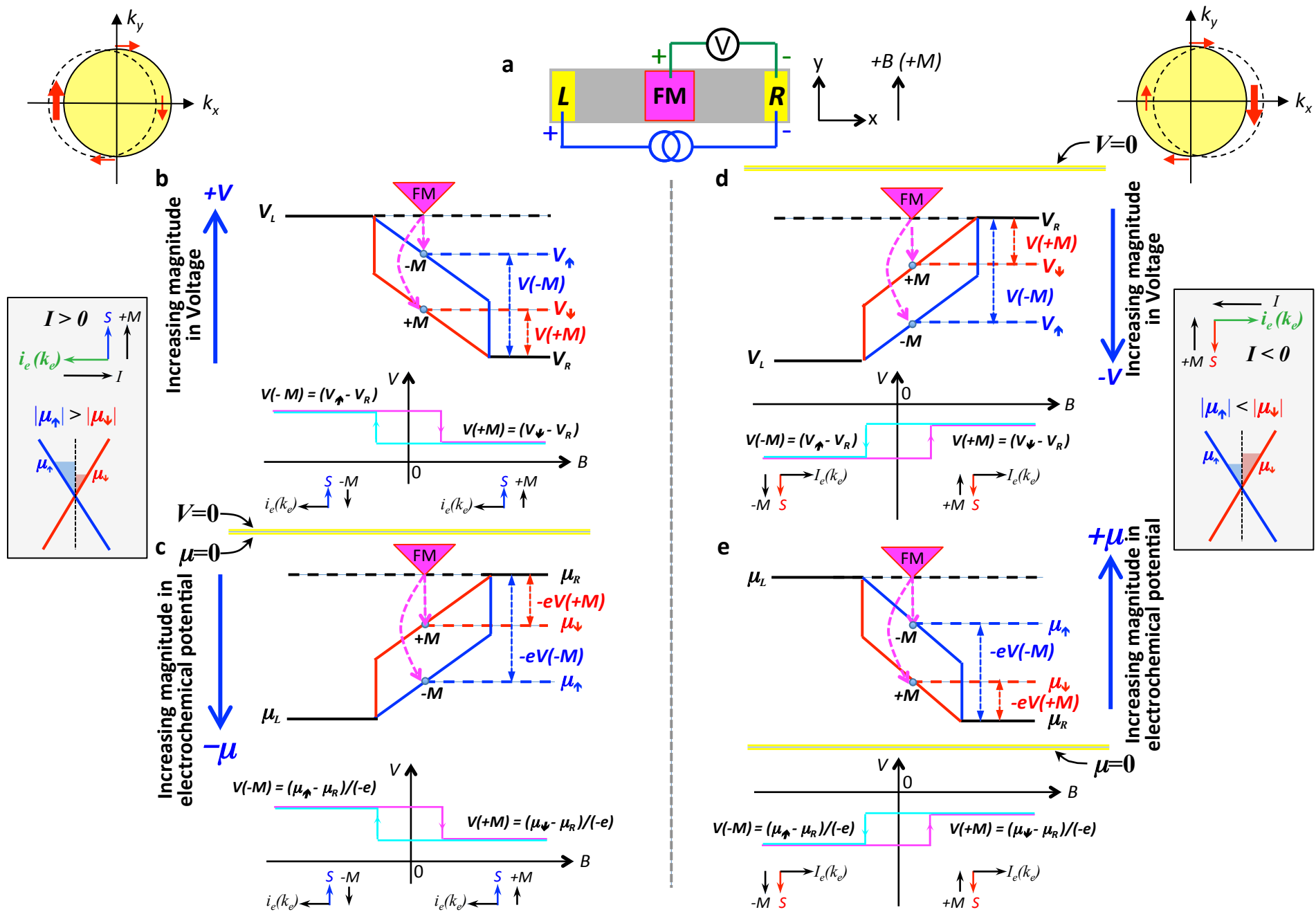

Li et al., Fig. 5